\documentclass[twocolumn,showpacs,amsmath,amssymb,superscriptaddress]{revtex4}

\usepackage{graphicx}
\usepackage{dcolumn}
\usepackage{bm}%

\begin{document}

\title[]{Cylindrically Symmetric Vacuum Solutions in Higher Dimensional Brans-Dicke Theory}

\author{ Ahmet Baykal}
\email{abaykal@nigde.edu.tr}
\affiliation{Department of  Physics, Faculty of Science and Letters, Ni\u gde University, 51240 Ni\u gde, Turkey}

\author{Dilek K. \c{C}iftci}
\email{dkazici@nku.edu.tr}
\affiliation{Department of Physics, Faculty of Science and Letters, Nam\i k Kemal University, 59030
Tekirda\u g, Turkey}

\author{\"Ozg\"ur Delice}
\email{odelice@gursey.gov.tr}
\affiliation{Feza G\"ursey Institute, P. K. 6  \c Cengelk\" oy, 34684 Istanbul, Turkey}
\affiliation{Department of Physics, Faculty of Science and Letters, Nam\i k Kemal University, 59030
Tekirda\u g, Turkey}

\date{\today}

\begin{abstract}
Higher dimensional, static, cylindrically symmetric vacuum solutions with and without a
cosmological constant in the Brans-Dicke theory are presented.
We show that for a negative cosmological constant and for  specific values of the parameters,
a particular subclass of these solutions include higher dimensional
topological black hole-type solutions with a flat horizon topology. We briefly extend our discussion to  stationary vacuum and $\Lambda-$vacuum solutions.
\end{abstract}
\maketitle

\section{Introduction}
\label{intro}
After the pioneering works of Kaluza and Klein \cite{Kaluza},
the exploration of higher dimensional gravity theories and their solutions played an important role in  physics.
They showed, in  particular, that adding an extra coordinate and  a scalar field, four dimensional gravity and electromagnetism can be unified upon compactification of the extra dimension on $S^1$.
Although this theory can be regarded  as speculative,
the underlying ideas inspired modern and more complicated theories, such as superstring theory \cite{String}.
The dilaton field in these theories behaves like  Brans-Dicke scalar field.

Brans-Dicke (BD) theory \cite{brans-dicke,Faraoni,brans}
is a  possible modification of Einstein's General Theory of Relativity (GR)
such that gravity is described by a metric geometry and, in addition, a scalar field.
The scalar field plays the role of a variable gravitational ``coupling constant''.
One of the original reasons where the theory was put forward was to attempt to incorporate the Machian ideas into the theory of gravitation.
In  BD theory, the matter fields   couple to the curvature as in the GR theory but not to the BD scalar field directly so that
energy momentum of matter fields are covariantly constant in the Jordan frame.
Moreover, since BD theory is a metric theory, weak equivalence principle holds in this frame.
It  passes solar system tests for large values of its free parameter $\omega$, i. e., for $\omega>40000$ \cite{Bertotti}.
It is also known that, metric $f(R)$ (modified) theories of gravity,  which have received a lot of attention recently, is dynamically equivalent to various versions of BD theory
with  an additional  potential term for the scalar field and particular values of the parameter $\omega$ \cite{SotiriouFaraoni}.
  Due to these and as well as other features of interest,
it is worthwhile to study its exact solutions
in comparison with GR counterparts in arbitrary number of dimensions.

Cylindrically symmetric solutions are one of the simplest and the most studied solutions of
gravity theories since
these solutions can be  considered as approximations to
non-spherical and elongated matter and field configurations \cite{SKMHH}.
For example, the cylindrical symmetry is adopted in the study of gravitational waves,
cosmological models, gravitational collapse of non-spherical matter distributions,
and in applications to the numerical GR as well.

Cylindrically symmetric, static vacuum solution of general relativity were presented by Levi-Civita long ago \cite{levicivita}.
This solution has two independent parameters \cite{Wang},
one of which is related to the mass per unit length
of the source whereas the other  is related to the
global conicity of the solution. An  important subclass of Levi-Civita (LC)
solution is  locally flat but globally conical cylindrically symmetric solution generated by a
 static, straight,  gauge cosmic string \cite{cosmicstring}. Due to the important gravitational and
 cosmological implications of cosmic strings \cite{Vilenkin,Hindmarsh}, this solution received a lot
 of interest in recent years. In four dimensional BD theory, the corresponding
 static cylindrical vacuum solutions were  presented in \cite{DahiaRomero,Arazi}, whereas the higher
 dimensional generalization of LC solution in GR is presented recently \cite{DeLeon}.

 In the presence of a cosmological constant $\Lambda$,
the corresponding solutions with  cylindrical symmetry was presented  much later by Linet \cite{Linet} and
 Tian \cite{tian}.
It is important to note that, for negative $\Lambda$ and definite choice of parameters \cite{dasilva},
this solution includes static topological black hole with a flat horizon topology as a special case
by an appropriate coordinate transformation \cite{Lemos,CaiZhang}.
For a review of topological black holes, see, e.g.,  \cite{Birmingham}.
As a possible source generating these $\Lambda-$vacuum solutions, thin shells were constructed in \cite{Zofka}.
The corresponding $\Lambda$-vacuum solutions for BD theory was discussed in \cite{Delice1}
and exact solutions corresponding to thick cosmic strings having a phenomenological
equation of state that matches smoothly to these solutions was presented in \cite{Delice2}.
The higher dimensional GR  generalizations of these solutions have recently been presented in \cite{BTekin1,BTekin2}.

In this paper, we present cylindrically symmetric,
static vacuum and $\Lambda$-vacuum solutions, generalizing the Levi-Civita's \cite{levicivita} and
 Linet-Tian's solution \cite{Linet,tian} to higher dimensional BD theory. We also present corresponding stationary (rotating) solutions for both cases in closed form.
 In our work we prefer to use Jordan frame,
however corresponding solutions in Einsten-scalar and dilaton gravity can be obtained by suitable conformal transformations.

\section{Field Equations}
\label{field}
The  vacuum field
 equations of BD theory with a cosmological constant $\Lambda$, in the Jordan frame can be derived from the action
\begin{eqnarray}\label{BDaction}
S&=&\int d^nx
\sqrt{|g|}\left\{\phi(-R+2\Lambda)+\frac{\omega}{\phi}\,g^{\mu\nu}\,\partial_\mu\phi
\,\partial_\nu \phi\right\}
\end{eqnarray}
in $n$ space-time dimensions.
The equations for the metric and the scalar field $\phi$ that follow from (\ref{BDaction}) can be written
in a form appropriate for study of BD $\Lambda$-vacuum as
\begin{eqnarray}
&&R_{\mu\nu}
=
(\omega+1)\, }{\tilde{\Lambda}\, g_{\mu\nu}
+
\frac{1}{\phi}\big(\phi_{,\mu;\nu}+\frac{\omega}{\phi}\phi_{,\mu} \phi_{,\nu} \big),
\phantom{aaa} \label{BDReq} \\
&&\Box\,\phi+\tilde{\Lambda}\, \phi=0,\label{BDFeq}
\end{eqnarray}
respectively.
Here $R_{\mu\nu}$ is the Ricci tensor, the constant
$\tilde{\Lambda}$ is related to the cosmological constant
and  the  BD parameter $\omega$ as
\begin{equation}
\tilde{\Lambda}
=
\frac{2\Lambda}{(n-1)+\omega(n-2)}.
\end{equation}
$\Box$ in (\ref{BDFeq}) denotes the Laplace-Beltrami operator in $n$ dimensional  space-time.

For the metric ansatz, we  adopt Gaussian normal type coordinates for  $n=4+d$ dimensional, static, cylindrically symmetric diagonal metric:
\begin{eqnarray}\label{metric}
ds^2&=&g_{\mu\nu}dx^\mu dx^\nu=dr^2+g_{ij}dx^i dx^j,
\\
 (g_{ij})&=&\mbox{diag}(-g_0^2,g_2^2,\ldots,g_{(n-1)}^2),
\end{eqnarray}
where  indices $\mu ,\nu$ cover all coordinate range,
whereas the indices $i,j$ cover range of the coordinates on the submanifold with $dr=0$.
All the metric functions $g_i$ depend only on the radial coordinate $r$.
In general, without any further symmetry assumption,  this metrical ansatz have at least $(n-1)$ number of commuting and hypersurface
orthogonal Killing vectors
$\xi^\nu_{(\mu)}=\delta^{\nu}_{\mu}, \mu\neq 1$. We shall also assume $\phi=\phi(r)$.
The function defined in terms of the field functions as  $W(r)=\phi\Pi_{i}g_i=\phi g_0 g_2\ldots g_{(n-1)}=\phi\sqrt{|g|}$
will be useful for the discussion below.

With the above assumptions,  the BD field equations (\ref{BDReq}) and (\ref{BDFeq}) lead to
the system of differential equations
\begin{eqnarray}
&&
\left(\frac{g_i'}{g_i}W \right)'=-(1+\omega)\tilde{\Lambda} W,\label{fei}
\\
&&
\left( \frac{W'}{W}\right)^2-\sum_{i}\left(\frac{g'_i}{g_i}\right)^2-\left(1+\omega \right)\left(\frac{\phi'}{\phi}\right)^2 =
-2\Lambda,\ \ \label{Weq2}
\\
&&\mbox{and}\nonumber
\\
&&
\left(\frac{\phi'}{\phi}W\right)'=-\tilde{\Lambda}W,\label{fephi}
\end{eqnarray}
respectively.
Using  (\ref{fei}) and (\ref{fephi}) we obtain
\begin{equation}\label{Weq}
 W''+\tilde{\Lambda} \left[(n-1)\omega+n\right]W=0.
\end{equation}
and that the function $W$ then satisfies
\begin{equation}
W'^2=k^2-\alpha W^2,\quad \alpha= \tilde{\Lambda} \left[(n-1)\omega+n\right]\label{Weq1},
\end{equation}
where $k$ is an integration constant.

Note that, the form of the field equations (\ref{fei})-(\ref{fephi}) are similar to four dimensional GR \cite{Linet}
and also to the BD \cite{Delice1,Petzold} cases.
The character of $W(r)$ which solves (\ref{Weq}) depends on
the numerical value of the constant $\alpha$,
which is determined by the numerical values of the parameters $\Lambda, \omega$ and $n$.

With the conformal transformations defined  by
\begin{equation}
\bar{g}_{\mu\nu}
=
g_{\mu\nu}\left(\frac{\phi}{\phi_0}\right)^{\frac{2}{n-1}},
\quad
\bar{\phi}
=
\left(\omega+\frac{n-1}{n-2} \right)^{\frac{1}{2}}
\ln\phi,
\end{equation}
$\phi_0$ being a constant,
the action (\ref{BDaction}) goes over to the Einstein frame with minimally coupled massless
scalar field and with an appropriately-scaled cosmological constant.
A subsequent conformal transformation defined by
\begin{equation}
\tilde{g}_{\mu\nu}
=
\bar{g}_{\mu\nu}\,e^{-\frac{4\phi}{n-2}},
\end{equation}
brings the action in the Einstein frame (in the barred variables)
to  so-called string frame where the scalar field now becomes dilaton field.

In the next section  we consider $\Lambda=0$ vacuum  solutions
which require $\alpha=0$ and $\omega\neq-\frac{n}{n-1}$. In the fourth section we present
$\Lambda$-vacuum solutions which require  either  $\alpha=0$ and $\omega=-\frac{n}{n-1}$  or $\alpha\neq 0$.
Finally we briefly extend our discussion to stationary generalizations of these solutions.

\section{Vacuum ($\Lambda=0$) Solutions}
\label{solutions1}
In this section we consider the case where the cosmological constant vanishes, $\Lambda=0$, with $\omega$ being arbitrary. For this case  Eqn.
(\ref{Weq}) becomes $W''=0$ and has   solution $W=\gamma\, r+\beta$ where $\gamma,\beta$ are integration constants.
It turns out that, without loss of generality, it is possible to choose $W=r$ by taking $\gamma=1$, $\beta=0$.
 The solution is therefore given by the functions
\begin{eqnarray}\label{solkas}
 g_i=A_i r^{a_i}, \quad \phi=A_\phi r^{a_\phi},
\end{eqnarray}
where the constant exponents satisfy
\begin{eqnarray}\label{kasnerrelations}
 \sum_i a_i+a_\phi=\sum_i a_i^2+(1+\omega)a_\phi^2=1.
\end{eqnarray}
This solution corresponds to higher dimensional generalization of the Levi-Civita space-time
\cite{levicivita} in BD theory and its four dimensional BD version can be found in \cite{DahiaRomero}.
The solution has  $n$ number of Kasner-like parameters $a_i,a_\phi$.
However, since there are two constraint equations, namely Eqns. (\ref{kasnerrelations}), only $n-2$ of them are independent.
We also have the freedom to set some of the constants $A_i$ to unity if the coordinate corresponding to
these constants are not an angular coordinate.
For the angular coordinate, however, the corresponding  constants is related to the global conicity of the space-time
and cannot be removed without changing the period of the angular coordinate.

This solution shares many physical properties with its four-dimensional counterpart.
For example, it is singular on $r=0$ since the Kretschmann scalar $\mathcal{K}=R_{\mu\nu\lambda\kappa}R^{\mu\nu\lambda\kappa}$,
diverges; as easily seen by  its expression,  for example, in $6$-dimensions:
\begin{equation}
\mathcal{K}\sim \frac{1}{r^4}\left\{\sum_i \left([a_i(1-a_i)]^2 -\tfrac{1}{2}a_i^2\right)+\tfrac{1}{2}\Big(\sum_i a_i\Big)^2\right\}.
\end{equation}
 The only exception  to this is  the particular
case where $a_\phi=0$ and all of the exponents $a_i$ vanish except one of them is assumed to be equal to unity.
This is nothing but the special case of (flat)  higher dimensional GR limit of our solution
and for  $n=4$ this solution describes the exterior field of gauge cosmic strings \cite{cosmicstring,Vilenkin}.

For $\Lambda=0$ case, the gravitational
force acting on a test particle  at rest is proportional to $\sim -a_0/r^{2a_0+1}.$
As in the four dimensional GR case \cite{Wang,Arik}, the direction of force depends on the sign of $a_0$.
When $a_0$ is positive the force is attractive whereas it is repulsive for negative $a_0$ implying
that the source generating such gravitational field has positive (negative) energy density  for positive (negative) $a_0$.

For the purpose of comparison with the familiar forms of the solution,
we first identify  the axial and the angular cylindrical coordinates as $x^2=z$, $x^3=\theta$.
We also label the extra coordinates by the capital Latin   letters, $I,J=(5, 6, \ldots, n)$.
Then, the solution  can be written in a more conventional form by incorporating the
Kasner relations (\ref{kasnerrelations})  into the solution by  the following coordinate transformation and redefinitions
\begin{eqnarray}
r&=& \rho^{\Sigma},\quad a_0=\frac{2\sigma}{\Sigma},\quad a_2=1-\frac{1}{\Sigma},\\
a_3&=&\frac{1-(2\sigma+p+\gamma_\phi)}{\Sigma},\quad a_I=\frac{\gamma_I}{\Sigma},\quad a_\phi=\frac{\gamma_\phi}{\Sigma}.
\end{eqnarray}
These yield the metric and the scalar fields as
\begin{eqnarray}
ds^2&=&-\rho^{4\sigma}dt^2+\rho^{2(\Sigma-1)}(dr^2+dz^2)\\&&+\rho^{2(1-2\sigma-p-\gamma_\phi)}\beta^2d\phi^2
+\sum_{I=5}^{n} \rho^{2\gamma_I}(dx^I)^2, \nonumber
\\
\quad \phi&=&\phi_0\rho^{\gamma_\phi},
\end{eqnarray}
respectively, where the constant exponents are now given by
\begin{eqnarray}
\Sigma&=&1-p+\frac{p^2+q^2}{2}+(2\sigma+\gamma_\phi)(p+\gamma_\phi-1)
+4\sigma^2 \nonumber\\&&+\frac{\omega}{2}\gamma_\phi^2,
\quad p=\sum_I \gamma_I,\quad q^2=\sum_I \gamma_I^2.\quad
\end{eqnarray}
Here, we re-scaled the ignorable coordinates for clarity.
Note that, this solution reduces to the corresponding
GR solution \cite{levicivita,DeLeon}
in the limit $\gamma_\phi\rightarrow 0$, $\sqrt{\omega}\rightarrow \infty$ such that $\gamma_\phi \sqrt{\omega}\rightarrow 0$.

\section{Vacuum Solutions with Cosmological Constant}

\subsection{General Case }
General class of solutions for the $\Lambda \neq 0$ case can be obtained for $\alpha \neq 0$. This requires
both $\Lambda \neq 0$ and $\omega \neq -\frac{n}{n-1}$.
Hence, the vacuum solution with a cosmological constant, which  can be considered as generalization of Linet-Tian solution \cite{Linet,tian} belongs to this class.
For this case, using Eqns.  (\ref{fei}) and (\ref{Weq}), the field equations can further be integrated to get
\begin{eqnarray}
\frac{g'_i}{g_i}&=&\frac{\omega+1}{n+(n-1)\omega}\frac{W'}{W}+\frac{kc_i}{W}\label{fei2}, \\
\frac{\phi'}{\phi}&=&\frac{1}{n+(n-1)\omega}\frac{W'}{W}+\frac{kc_\phi}{W}\label{feP},
\end{eqnarray}
where $c_i$ and $c_\phi$ are constants. These constants,  because of  the equations (\ref{Weq2}) and (\ref{Weq1}), should satisfy
the following constraint equations:
\begin{eqnarray}
&&\sum_{i=0}^{n-1}c_i+c_\phi=0,\\
&&\sum_{i=0}^{n-1}c_i^2+\left(1+\omega \right)c_n^2=\frac{n-1+\left(n-2 \right)\omega }{n+\left(n-1 \right)\omega}.
\end{eqnarray}

Eqn. (\ref{Weq}), depending on the sign of $\alpha$, has two types of well-known solutions.  These are given by
\begin{eqnarray}
W
&=&
\gamma \sin \left(\sqrt{\alpha}\,r\right)+\delta\cos\left(\sqrt{\alpha}\,r\right),\quad\quad\quad  \alpha>0,
\\
W
&=&
\gamma \sinh\left(\sqrt{|\alpha|}\, r \right)+\delta\cosh\left(\sqrt{|\alpha|}\,r\right),   \alpha<0.
\end{eqnarray}
Note that cylindrical symmetry requires $\delta=0$ and we have the freedom to take $\gamma=1$.
Now, the remaining functions of the solution can be found. Using eqns. (\ref{fei2}) and (\ref{feP})
we find the following field functions
\begin{eqnarray}
g_i
=
C_i \left(\tan{\tfrac{1}{2}}\sqrt{\alpha}\,r\right) ^{c_i}\left( \sin{\sqrt{\alpha}\,r}\right)^{\frac{\omega+1}{n+\left(n-1 \right)\omega }},
\\
\phi
=
C_\phi \left(\tan\tfrac{1}{2}\sqrt{\alpha}\,r\right) ^{c_\phi}\left( \sin{\sqrt{\alpha}\,r}\right)^{\frac{1}{n+\left(n-1 \right)\omega }},
\end{eqnarray}
for $\alpha>0$, and
\begin{eqnarray}
g_i
=
C_i \left(\tanh{\tfrac{1}{2}}\sqrt{|\alpha|}\,r\right)^{c_i}\left( \sinh{\sqrt{|\alpha|}\,r}\right)^{\frac{\omega+1}{n+\left(n-1 \right)\omega }},\ \
\\
\phi
=
C_\phi \left(\tanh{\tfrac{1}{2}}\sqrt{|\alpha|}\,r\right)^{c_\phi}\left( \sinh{\sqrt{|\alpha|}\,r}\right)^{\frac{1}{n+\left(n-1 \right)\omega }},\ \
\end{eqnarray}
for $\alpha<0$.

Similar to the GR  \cite{Linet,tian} as well as BD \cite{Delice1} cases in four dimensions,
the curvature invariants of the solutions are singular at $r=0$ and at $r=N\pi/\sqrt{\alpha}$
 for  $\alpha>0$ where $N$ is an odd integer,  whereas they are singular only at $r=0$ for $\alpha<0$. Hence, for negative $\alpha$ spacetime is similar to vacuum case where the only singularity is at the symmetry axis, which can be avoided for both signs of $\alpha$ if one can put an interior regular source occupying the symmetry axis and smoothly matching the exterior $\Lambda-$vacuum region. Positive $\alpha$ case is more problematic however, since there are infinitely many singularities at  $r=N \pi/\sqrt{\alpha}$. This can be remedied by  assuming that
 this case describes a closed toroidal spacetime with topology $R^{n-2}\times S^2$ as suggested in \cite{tian}.   Then, the radial coordinate is to be transformed to $r=\chi+\chi_0$ where $\chi_0=\pi/\sqrt{\alpha}$ and  the new coordinate $\chi$ has  range $-\chi_0\le\chi\le \chi_0$.

In order to discuss the asymptotical form of the solutions, with the help of the properties of trigonometric functions, we rearrange the metric functions given above as follows
\begin{eqnarray}
 g_i&=&C_i \cos^{\frac{2(\omega+1)}{n+(n-1)\omega}}(\frac{\sqrt{\alpha}\,r}{2})\tan(\frac{\sqrt{\alpha}\,r}{2})^{a_i},\\
\phi&=&C_\phi \cos^{\frac{2}{n+(n-1)\omega}}(\frac{\sqrt{\alpha}r}{2})\tan(\frac{\sqrt{\alpha}\,r}{2})^{a_\phi},
\end{eqnarray}
 for $\alpha>0$  and for $\alpha<0$ trigonometric functions replacing with corresponding hyperbolical functions. Here the parameters $a_i,a_\phi$, are  given by
\begin{eqnarray}
 a_i=c_i+\frac{1+\omega}{n+(n-1)\omega},\quad a_\phi=c_\phi+\frac{1}{n+(n-1)\omega},\  \ \
\end{eqnarray}
and they satisfy the relations (\ref{kasnerrelations}).

Near $r\approx 0$, the solutions, for both signs of $\alpha$, reduce to vacuum Levi-Civita-BD solution (\ref{solkas}):
\begin{eqnarray}
 g_i(r)_{r\rightarrow 0}\approx r^{a_i}, \quad \phi(r)_{r\rightarrow 0}\approx r^{a_\phi}.
\end{eqnarray}
For positive $\alpha$,  the solutions  do not approach to de-Sitter spacetime in the $r\rightarrow \infty$ limit.
However, for $\alpha<0$, the metric reduces to
\begin{eqnarray}
g_i(r)_{r\rightarrow \infty}
\approx dr^2+e^{\frac{2(\omega+1)}{n+(n-1)\omega}\sqrt{|\alpha|} r}\Big(\sum_{i\neq0}^n(dx^{i})^2-dt^2\Big),\ \ \ \ \ \
\end{eqnarray}
which is actually a spacetime with a constant negative curvature. Hence, the negative $\alpha$ space-time always approaches to anti de Sitter (AdS) space-time at radial infinity. Note that in the limit $r \rightarrow \infty$, for this case, the metric becomes independent of the cylindrical Kasner parameters.

\subsubsection{Relation with Topological Black Holes}
Let us consider a subclass of solutions for $\alpha<0$.
First, consider the 
coordinate transformation defined by
\begin{eqnarray}
 r
 &=&
 \frac{2}{\sqrt{|\alpha|}}
 \cosh^{-1}\left[ \sqrt{\Omega} \left(\frac{\rho}{\rho_0}\right)^{ \frac{n+(n-1)\omega}{2(1+\omega)} }  \right],
 \\
\Omega&=&\frac{(1+\omega)^2 |\tilde{\Lambda|}\rho_0^2}{n+(n-1)\omega},\\[.2cm]
\nonumber
\end{eqnarray}
($\rho_0$ constant)
together with  the special choice of  the Kasner parameters
\begin{eqnarray}
&&c_2=c_3=c_I=c_\phi=\frac{-(1+\omega)}{n+(n-1)\omega}, \\&& c_0=\frac{n-1+(n-2)\omega}{n+(n-1)\omega},
\end{eqnarray}
which brings this special case of general BD vacuum solution with $\alpha<0$  to the
convenient form
\begin{eqnarray}
 ds^2
 &=&
 - \left(\frac{\rho}{\rho_0}\right)^2\left[\Omega-\left(\frac{\rho_0}{\rho}\right)^{\frac{n+(n-1)\omega}{1+\omega}} \right]dt^2
\nonumber   \\
&+&
\left(\frac{\rho}{\rho_0}\right)^{-2} \left[\Omega-\left(\frac{\rho_0}{\rho}\right)^{\frac{n+(n-1)\omega}{1+\omega}} \right]^{-1}d\rho^2
\nonumber  \\
&+&
\left(\frac{\rho}{\rho_0}\right)^2 \bigg[dz^2 + C_\phi^{'2} d\phi^2 +\sum_I(dx^I)^2\bigg],\label{bh}\\
\phi&=& \phi_0 \left[(1+\omega)\rho \right]^{\frac{1}{1+\omega}}.
\end{eqnarray}
Here, the time and other ignorable coordinates are scaled by  constants. To avoid a conical singularity, one has to set $C_\phi'/\rho_0$ to unity.

This solution corresponds to an interesting subclass of general $\Lambda-$vacuum BD  solutions  with cylindrical symmetry. It is
presented in \cite{Dehghani} and its thermodynamical and
 other properties are discussed in
\cite{Dehghani} in the study of dilaton gravity with a dilaton potential term.
The solution (\ref{bh}) is the higher dimensional generalization of the solution given by Cai et al. \cite{cai}
corresponding to  black hole-type solutions with flat (cylindrical, plane or toroidal) horizon topology in the dilaton gravity,
whose GR version were previously  studied  in four \cite{Lemos,CaiZhang} and higher dimensions \cite{Awad} .
In fact, this type of solutions in GR has also been identified with the AdS solitons \cite{BTekin1,BTekin2,Horowitz}.
As in the  corresponding case in GR, the presence of negative cosmological constant  is essential for these black hole-type solutions.
Moreover, by omitting the coordinates of the extra dimensions, the solution (\ref{bh}) goes over to the corresponding GR case
in the  limit $\omega\mapsto\infty$.

\subsection{Solutions with $\alpha=0$}
The class of solutions for $\alpha=0$ with $\Lambda\neq 0$  is given by the  particular value of the BD parameter
$\omega=-\frac{n}{n-1}$. For this class of solutions, the field functions are given by
\begin{eqnarray}
 g_i=B_i\, r^{b_i}\, e^{\Lambda r^2/2},\quad \phi=B_\phi\, r^{b_\phi}\,e^{(1-n)\Lambda r^2/2},
\end{eqnarray}
where the constants are  to satisfy
\begin{eqnarray}
 \sum_i b_i+b_\phi=\sum_i b_i^2-\frac{1}{(n-1)}b_\phi^2=1.
\end{eqnarray}
Note that this solution has no GR limit, unless $\Lambda=0,b_\phi=0$ which renders the  BD scalar function a constant.

\section{Stationary Solutions}

In GR, cylindrical stationary (rotating) solutions were presented for vacuum by Lewis\cite{Lewis} and $\Lambda-$vacuum by Krasinski\cite{Krasinski} and Santos\cite{Santos}. These solutions can be obtained by using  globally forbidden coordinate transformations from the corresponding static solutions, i.e. coordinate transformations which mixes time and angular coordinates \cite{Stachel1,Lemos}.
This type of transformation yields locally equivalent but globally different solutions. Now suppose that for the metric we considered (\ref{metric}), the spacetime has  $R^{n-1}\times S^1$ topology and the coordinate $x^3$ is an angular coordinate. 
Then,
as in the previous cases, we can obtain stationary (rotating) cylindrical solutions applying the following coordinate transformations
\begin{equation}\label{transstat}
t=t_0 T+  x_0 \Phi, \quad x^3=t_1 T+x_1 \Phi,
\end{equation}
where $t_0,x_0,t_1,x_1$ are constants, and there is  freedom to rescale two of them to unity by redefining the coordinates $T$ and $\Phi$.
Consequently, the
resulting metric takes the form:
\begin{eqnarray}
ds^2&=&-FdT^2+2K dT d\Phi+L d\Phi^2+dr^2\nonumber \\
&&   +\sum_{i\neq 0,3}^n (g_i dx^i)^2
\end{eqnarray}
where
\begin{eqnarray}
 &&F=t_0^2 g_0^2-t_1^2 g_3^2,\quad K=t_1 x_1 g_3^2-t_0 x_0 g_0^2,\nonumber \\ &&L=x_1^2 g_3^2-x_0^2 g_0^2.
\end{eqnarray}

Since the transformation is not  permitted globally  and the resulting metric is not equivalent to static metric, we can  label the new coordinate $\Phi$ as angular coordinate with range $[0,2\pi]$. One can always locally transform the new metric into a static one by considering the inverse transformations of (\ref{transstat}), but  since these transformations are not permitted globally, the static metrics can  fully be recovered only in the limit where the constants $t_1$ and $x_0$  vanish.

Using appropriate forms of the functions $g_i$ presented in previous sections, the above metric describes stationary vacuum or $\Lambda-$vacuum solutions of cylindrical
stationary solutions of higher dimensional BD theory or with taking suitable limits, GR theory.

Since static and stationary rotating solutions are equivalent locally, many of their  physical properties are also similar.  
 However, there is a crucial difference between static and stationary solutions. Cylindrically symmetric  rotating solutions can easily generate closed timelike curves (CTC's).  For the metric under consideration, if $\xi^\mu_{(\Phi)}$ is the Killing vector generating rotations about axis, then the spacetime is free of CTC's if $g_{\mu\nu}\xi^\mu_{(\Phi)}\xi^\nu_{(\Phi)}>0$ everywhere, which requires 
 $g_{0}^2>x_0^2/x_3^2\,g_3^2$. It is possible to have such configurations with appropriate choice of parameters. For example, for the vacuum, the case $a_0>a_3$ ($a_3>a_0$) is free of CTC's for large $r$ if the constants satisfy the relation $x_0^2>x_3^2$ ($x_3^2>x_0^2$).

 \section{Discussion}
 \label{discussion}

In this paper, various  solutions of Brans-Dicke field equations having cylindrical symmetry in higher dimensions are presented for
the cases of vacuum and  $\Lambda$-vacuum. The
solutions we have presented generalize the corresponding solutions for static vacuum \cite{levicivita} and $\Lambda$-vacuum
\cite{Linet} cases and stationary vacuum \cite{Lewis} and $\Lambda-$vacuum \cite{Krasinski,Santos} in GR.  We have showed in particular that, similar to corresponding to the GR case, a subclass
of higher dimensional $\Lambda$-vacuum solutions we present includes
a topological black hole-type solution for the specific choice of parameters of
the higher-dimensional BD theory. It is well-known that one of the ingredients of the low energy limit of string theories is   dilaton field that couples nonlinearly to gravity. Thus,  the solutions we have discussed can have relevance in the context of
such theories,
since they  represent the gravitational field of static or stationary rotating  neutral string-like objects in four and higher dimensions.

\end{document}